\documentstyle [aps,amssymb,prl,preprint]{revtex}
\draft
\begin{document}
\date{July 5, 1999}

\title{Non-stationary Characteristics of the instability in a
Single-mode Laser with Fiber Feedback}

\author{Tsung-Hsun Yang, Tsong-Hsin Lim, Jyh-Long Chern, and Kenju Otsuka$^*$}

\address{Nonlinear Science Group, Department of Physics, National Cheng
Kung University, Tainan 70101, Taiwan, Republic of China\\
$^*$Department of Applied Physics, Tokai University, 1117
Kitakaname Hiratsuka, Kanagawa 259-12, Japan}

\maketitle

\begin{abstract}
Chaotic bursts are observed in a single-mode microchip Nd:YVO$_4$
laser with fiber feedback. The physical characteristic of
the instability is a random switching between two different dynamical
states, {\it i.e.}, the noise-driven relaxation oscillation and the
chaotic spiking oscillation. As the feedback strength varied,
transition which features the strong interplay between two states
exhibits and the dynamical switching is found to be non-stationary at
the transition.

\end{abstract}

\vspace{2.0cm}


The complex dynamics in nonlinear systems with delayed feedback
which possesses infinite dimensions are of current interest in
various fields including physics, chemistry, biology, economy,
physiology, neurology, and optical systems \cite{1}. Especially,
instabilities of nonlinear optical resonators and lasers with
delayed feedback have attracted much attention in the past decade.
In history, the issue of chaotic instabilities in the output of
lasers which is subjected to external feedback was initiated by
the pioneering work of Lang and Kobayashi in 1980\cite{2}. They
demonstrated the dynamical instabilities in a semiconductor laser
with external feedback which features the sustaining relaxation
oscillations. They also confirmed theoretically that the dynamical
instabilities take place in the transition process where the
lasing frequency changes from one external cavity eigenmode to
another in a weak-coupling regime. Thereafter, three universal
transition routes to chaos, low-frequency fluctuations (LFF) and
coherence collapse have been observed in semiconductor lasers with
external optical feedback for different feedback strength and/or
delay time regions\cite{3}. To LFF and coherence collapse, there
still exists the open question to concern the role of noise
\cite{4}. Nowadays, another promising laser system for
investigating the instabilities in lasers with delayed feedback
would be the laser-diode-pumped microchip solid-state lasers which
have been widely used in the practical applications. They are
expected to exhibit extremely high-sensitive response to the
external feedback. The reason is that the cavity round-trip time
$\tau _{L}$ ($\varpropto \tau _{p}$: photon lifetime) compared
with the fluoresecence lifetimes $\tau $ is extremely short as
demonstrated in self-mixing laser Doppler velocimetries \cite{5}.
Generally, the lifetime ratio $K=\tau /\tau _{p}$ of solid lasers
ranges from 10$^{5}$ to 10$^{7}$, while $K\backsimeq 10^{3}$ in
laser diodes.  Furthermore, their characteristic frequencies are
of sub-MHz, thus the conventional measurement techniques can be
utilized easily.  Therefore, it will be much easier to study the
various instabilities in solid-state laser systems than in laser
diode systems. In fact, in the early experiment in 1979, Otsuka
observed various instabilities in a microchip LNP
(LiNdP$_{4}$O$_{12}$) solid-state laser subjected to external
feedback\cite{6}. However, the instability is only observed in the
region of two lasing modes. It is expected that the instability
can also occur due to the external cavity modes with small mode
spacing. The mechanism is that a random intensity fluctuation in
each mode will result in mode-dependent random fluctuation in
phase shift\cite{6}. Hence, multimode oscillation and intrinsic
mode-partition noise as well as frequency dependent nonlinear
refractive index are the dynamical origins to cause the chaotic
bursts in the presence of fiber\cite{6}. Here we report an
experimental result of a diode-pumped microchip Nd:YVO$_{4}$
(yttrium orthovanadate) solid state laser, in which mode-partition
noise is not essential, but the intrinsic phase fluctuation may be
dominant around the onset of instability. Furthermore, it will be
shown that a novel non-stationary characteristic is inherent in
the instability due to the strong coupling between two different
dynamical states at transition. This non-stationary feature should
be unique for the characterization of generic dynamical systems.

 In our experiment, a diode-pumped microchip Nd:YVO$_4$
laser operated at {\it single-mode} regime is employed and a
compound cavity is formed with single-mode fiber feedback. The
laser diode (LD) and Nd:YVO$_{4}$ (1mm thick, 1$\%$ Nd$^{3+}$
doped, and the output coupling is $5\pm2\%$ at 1064nm) are
available from CASIX, Inc. The laser crystal (Nd:YVO$_{4}$) is
inserted into a 2mm-thick copper mount and the temperature is
controlled at 25$^{o}$C by a temperature controller (ILX,
LDT-5910B). The pumping beam (with wavelength at
$\lambda_{p}$=808nm) from LD, which is also
temperature-controlled, is focused onto the laser crystal with a
GRIN lens (0.22 pitch). The pumping threshold is around 300 mA. We
also use a noise filter (ILX 320) to eliminate the pumping noise
caused by the LD current driver (ILX, LDC-3744) and an
interference filter (60$\%$ transmission at 1064nm and
zero-transmission for the rest) to reduce the influence of pumping
light on detection.  In the entire pumping domain a
$\pi$-polarized TEM$_{00}$ mode of laser output was observed. We
used a 10m single-mode fiber (3M F-SY) as the feedback loop. The
feedback beam is monitored and no significant polarization has
been found. Because of the reduction of the lasing threshold
(about 1-2 mA less) the feedback strength is estimated to be below
1$\%$.  To further control the feedback strength, before the light
entering the fiber, a rotatable polarizer (New Focus 5525) is
utilized. In the measurement, a multi-wavelength meter (HP 86120B)
is employed to monitor the variation of lasing mode and the lasing
wavelength is 1064.245 nm.  The lasing eigenmode frequency of the
compound cavity is determined by the frequency arrangement of the
Nd:YVO$_{4}$ laser cavity mode (mode spacing 0.25 nm; i.e., $\sim$
60 GHz) and the external cavity (fiber) modes for which the number
of modes is very large (the mode spacing of external cavity is
around 10 MHz).  We also utilize low-noise detectors (New Focus
1611; bandwidth 1GHz) for the detection of laser output. Both ac
and dc ports of the detectors are connected to a transient
oscilloscope (HP54542C) for data acquisition in the temporal
domain. Meanwhile, a rf-spectrum analyzer (HP8591E) is employed to
monitor the behavior of laser output in the rf-spectrum domain.

For later identification, we first show the result of a typical ac
time series for the case of free-running and its corresponding
rf-spectrum in Fig.\ref{fig:free-running}. Relaxation oscillation
occurs around 1.6 MHz and its harmonics can be easily
characterized. As the strength of feedback is increased, chaotic
bursting occurs and the dominant frequency was shifted to a lower
value (around 1.0 MHz) with broadened-linewidth. Typical time
series of the chaotic bursting is shown in
Fig.\ref{fig:fiber-feedback}~(a). To explore the dynamics, we
employ a joint-time frequency analysis. The coincidence of
frequency characteristics between Fig.\ref{fig:free-running}~(b)
and Fig.\ref{fig:fiber-feedback}~(b) suggests that the
low-intensity level part of chaotic bursting, the regime {\rm I}
in Fig.\ref{fig:fiber-feedback}~(a), is a noise-driven relaxation
oscillation while the high-intensity level part, the regime {\rm
III} in Fig.\ref{fig:fiber-feedback}~(a), can be identified to be
chaos based on a singular value decomposition analysis\cite{7}. We
noted that the basic characteristics of the dynamical transition
between two states is frequency-broadening in the rf-spectrum as
shown in Fig.\ref{fig:fiber-feedback}~(c).  Nevertheless, the
lasing wavelength remains the same. That is no significant
line-width broadening or hopping has been seen.  Meanwhile, as
instability occurs, the signal-to-noise level of the lasing mode
decreases and features as a fast frequency-modulated (FM) laser
characteristics. This suggests that there is a FM noise in the
process of instability.

Next let us address the physical mechanism of coexistence of two
dynamical states.  As seen from the time series, the behavior of
peak power is a key factor for the dynamics. Since our system is
essentially a single-mode laser with weak feedback, the
Lang-Kobayashi model\cite{2} is still applicable such that the
photon density $S(t)$ follows
\begin{equation}
\frac{dS(t)}{dt}=K[(n(t)-1)S(t)+\epsilon n(t)]+2 \kappa
\sqrt{S(t)S(t-T)}cos\theta(t),
\label{eq:St}
\end{equation}
where $K$
is the time ratio between the population inversion lifetime and the
photon lifetime, $n(t)$ is the population inversion density (or carrier
density), $\epsilon$ is the spontaneous emission factor, $\kappa$ is
the feedback coupling strength, $T$ is the delay time, and $\theta$ is
the phase difference between the output and the feedback beams. The
peak power $S_{p}$, therefore, follows
\begin{equation}
S_{p}=\frac{-K\epsilon n_{p}(t)}{K(n_{p}(t)-1)+2\kappa
\sqrt{1+\frac{\Delta S}{S_{p}}}cos\theta_{p}},
\label{eq:Sp}
\end{equation}
where $\Delta S=S_{p}(t)-S_{p}(t-T)$ and the subscript
$p$ denotes the corresponding quantities evaluated at $S=S_{p}(t)$. The
role of $\Delta S$ is crucial. As the feedback coupling $\kappa$ is
almost zero, the statistics $S_{p}$ should simply follow the (on-off)
fluctuation of the population inversion as implied by the lasing
threshold factor, $n(t)-1$. With a nonzero-$\kappa$, the dynamics of
$S_{p}$ will be modified by the appearance of $\Delta S$ as well as the
phase term $\theta_{p}$.  Since the magnitude of the peak output is
directly measurable, a further investigation of the peak photon
intensity and the statistics of time-difference quantity ($\Delta S$)
will be fruitful. With oscilloscope, we repeatedly accumulate the time
series of laser output and pick up the discrete peak value
${S_{p}(n)}$, $n=1,2,...$ where $n$ denotes the $n$th peak. To have a
reliable probability, total $320,000$ peaks have been collected for
average at any specific rotation angle of polarizer. As the polarizer
is rotated, the feedback strength will be changed. Asymptotically, as
$\kappa\rightarrow 0$, the system features a free-running laser such
that the mean and the standard deviation are small. However, there is a
dramatic increasement in both of the mean and the standard deviation
around 41-42 degree of polarizer's angle which implies the onset of
transition. To further identify the transition, we evaluate the
probability distribution of peak powers at different feedback strength
(equivalent to the different polarizer's angle). In the regime of
noise-driven relaxation oscillation, the probability distribution of
peak power $P(S_{p})$ follows an exponential law which features a
shot-noise characteristics (as shown in Fig.\ref{fig:probability}~(a)).
On the other hand, in the case that chaotic bursting occurred, a tailed
probability distributions will be created. If we pay attention only on
the part of large intensity, by which the exponential distribution can
be neglected, a Gaussian distribution can be recognized as feedback
strength is further increased as shown in Fig.\ref{fig:probability}
(c). This shows that there are two dynamical states which follows
different statistics.  Dynamical transition from a simple exponential
distribution to a mixed distribution does occur as shown in
Fig.\ref{fig:probability}~(a)-(c).

What is the influence of the onset of such a mixed distribution?
By a joint probability analysis similar that used in \cite{8}, it
can be identified that there is still a strong overlapping between
the two probability distributions. This suggests that the
interplay between two states may be rather unique where the
statistics of time-difference should be crucial also as discussed
above. This also means that we should look at the dynamical
behavior of a $k$-step difference quantity,
\begin{equation}
\Delta S_{p}(k)=S_{p}(k+l)-S_{p}(l),
\label{eq:Sp(k)}
\end{equation}
where $S_{p}(l)$ is the $l$th. peak power. After the summation
over the whole range of $l$, the probability $P(\Delta S_{p}(k))$
specifies the {\it variation distribution} with $k$-step
difference. Consider that when the system is with noise-driven
relaxation oscillation, the variation distributions should be the
same no matter what is the value of $k$ since the switching
characteristic is stationary. The difference can be characterized
by a $\chi^{2}$ statistics which is defined as
\begin{equation}
\chi^{2}(j;k)=\sum_{i}^{M}\frac{(R_{i}-S_{i})^{2}}{(R_{i}+S_{i})},
\label{eq:chi2_j_k}
\end{equation}
where $R_{i}$ and $S_{i}$ are the probabilities of the $i$th
interval of $\Delta S_p$ for $P(\Delta S_{p}(k+j))$ and $P(\Delta
S_{p}(k))$, respectively.  The summation is carried out for all
intervals except $R_{i}=S_{i}=0$\cite{9}. For a fixed $j$, the
$\chi^{2}(j;k)$ indicates the {\it similarity} of the distribution
between the variations $\Delta S_{p}(k)$ and $\Delta S_{p}(k+j)$.
Foremost of all, the most crucial quantity is the {\it successive
change on similarity}, {\it i.e.}, $j=1$.  A large value in
$\chi^{2}(1;k)$ means that at time $k$ the dynamics has been
switched to a state which follows a dramatically different
statistical distribution and this suggests a strong interplay
between states. Furthermore, when the value of $\chi^{2}(1;k)$ is
wildly varied as $k$ moved a non-stationary dynamics will be
intrinsic in nature. The degree of non-stationarity of the whole
process can be quantified by an average quantity, i.e.
\begin{equation}
\chi^{2}(1)=\frac{1}{L}\sum_{k=1}^{L}\chi^{2}(1;k).
\label{eq:chi2_1}
\end{equation}
In principle, the correlation of variation distributions is
essentially stationary as $\chi^{2}(1) \approx 0$\cite{9}, i.e.,
there is no dramatic dynamical changes on the correlation of the
variation distributions in the range of average. Shortly, one can
explore the dynamical characteristics of switching with this
$\chi^{2}$ statistics. Let us next address our experimental
results.  As shown in Fig.\ref{fig:probability-association} (a)
where $L=200$, $\chi^2(1)$s are almost close to zero for the
relaxation oscillation regime as expected. The interesting point
is the appearance of high $\chi^2(1)$ at the medium strength of
feedback (around $40^\circ$ of polarizer's angle), which also
corresponds to the transition indicated by the probability
distribution, the mean, and the standard deviation. This shows
that the transition is associated with a non-stationary
characteristic and, thus, the successive change on similarity is
wild. It remains a surprise that for a larger feedback strength
(small polarizer's angle), $\chi^2(1)$ is still almost zero again.
This peculiar feature is due to the observation that an increase
of feedback strength causes an increase of staying times at
particular state and, as a result, the successive change on
similarity will become smooth. In more details,
Fig.\ref{fig:probability-association}~(b) presents the degree of
probability association, $\chi^2(1;k)$. For some particular
feedback strength, $\chi^2(1;k)$ never keeps near zero, which
suggests the instability has a strong non-stationary switching. On
the other hand, in the regime of large polarizer's angle
(noise-driven relaxation oscillation), the $\chi^2(1;k)$ is almost
zero (Fig.\ref{fig:probability-association}~(c)). Here, we would
like to emphasize that the transition indicated by a high
$\chi^{2}(1)$ is crucial and it reveals a rather amazing
characteristics: a {\it weak}-feedback induced instability can
even be associated with a {\it wild} and {\it non-stationary}
successive change on the similarity of variation probability
distributions.

Let us discuss the origin of such chaotic bursting and even
non-stationary characteristics in such an instability.  Referring
to Eq.(\ref{eq:St}) and Eq.(\ref{eq:Sp}), since the dc value of
the photon density $S(t)$ is never zero, the term $cos\theta(t)$
sometimes has to be zero such as to terminate the feedback
process. In such a way, the noise-driven relaxation oscillation
can be created with non-zero $\kappa$. This means that a fixed
phase difference ($\theta\approx\pi/2$) during the period of
relaxation oscillation has to be established. Therefore, the
persistence in fixed phase difference is essential for switching
back to the relaxation oscillation and should be critical for the
overlapping in joint-probability as well as the non-stationary
switching. In other words, the random phase fluctuation due to the
feedback effect is essential for the chaotic bursting behavior.
Hence, an inclusion of fast frequency-modulation characteristics
(FM noise) to the phase fluctuation is necessary in the process as
experimentally suggested from the measurement of the
multiwavelength meter. Indeed, under this condition, the
simulation of the single-mode Lang-Kobayashi model can reproduce
the chaotic bursting features reported above. Finally, we would
like to note that the similar instability has also been observed
with various fiber lengths (7m, 5m, 4m, and 3m) as well as
multimode fiber.  Moreover, in stead of polarizer, a N.D. filter
is also employed and similar results are concluded. The further
details as well as the analysis of $\chi^{2}(j)$ with large-$j$
will be reported elsewhere.

{\it Acknowledgment:} The work of NCKU group is partially
supported by the National Science Council, Taiwan, ROC under
project no. NSC88-2112-M-006-001.  JLC thank S.-L Hwong, J.-Y. Ko,
and A.-C. Hsu in helping the preparation of manuscript.

\begin{figure}
\caption{Typical (a) ac time series and (b) its corresponding
rf-spectrum for the case of free-running.}
\label{fig:free-running}
\end{figure}

\begin{figure}
\caption{(a) The time series of chaotic bursting behavior due to
the feedback effect and its corresponding spectra according to the
joint-time frequency analysis on (b) the regime \rm{I}, (c) the
regime \rm{II}, and (d) the regime \rm{III}, respectively.}
\label{fig:fiber-feedback}
\end{figure}

\begin{figure}
\caption{The probability distribution of peak powers at different
feedback strengths indicated by the degrees of rotation of the
polarizer at (a) $54^\circ$, (b) $41^\circ$, and (c) $0^\circ$.}
\label{fig:probability}
\end{figure}

\begin{figure}
\caption{(a) The average probability association $\chi^2(1)$ at
different feedback strengths (different angles of the polarizer) and
the probability association $\chi^{2}(1;k)$ versus different $k$ at (b)
$40^\circ$ and (c) $54^\circ$ of polarizer's angles.}
\label{fig:probability-association}
\end{figure}

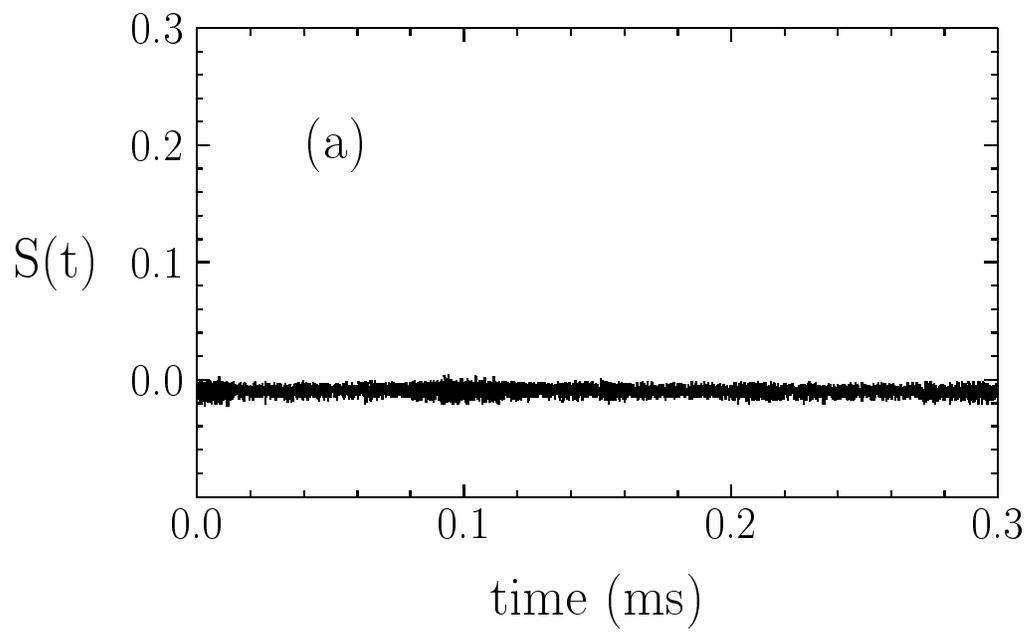
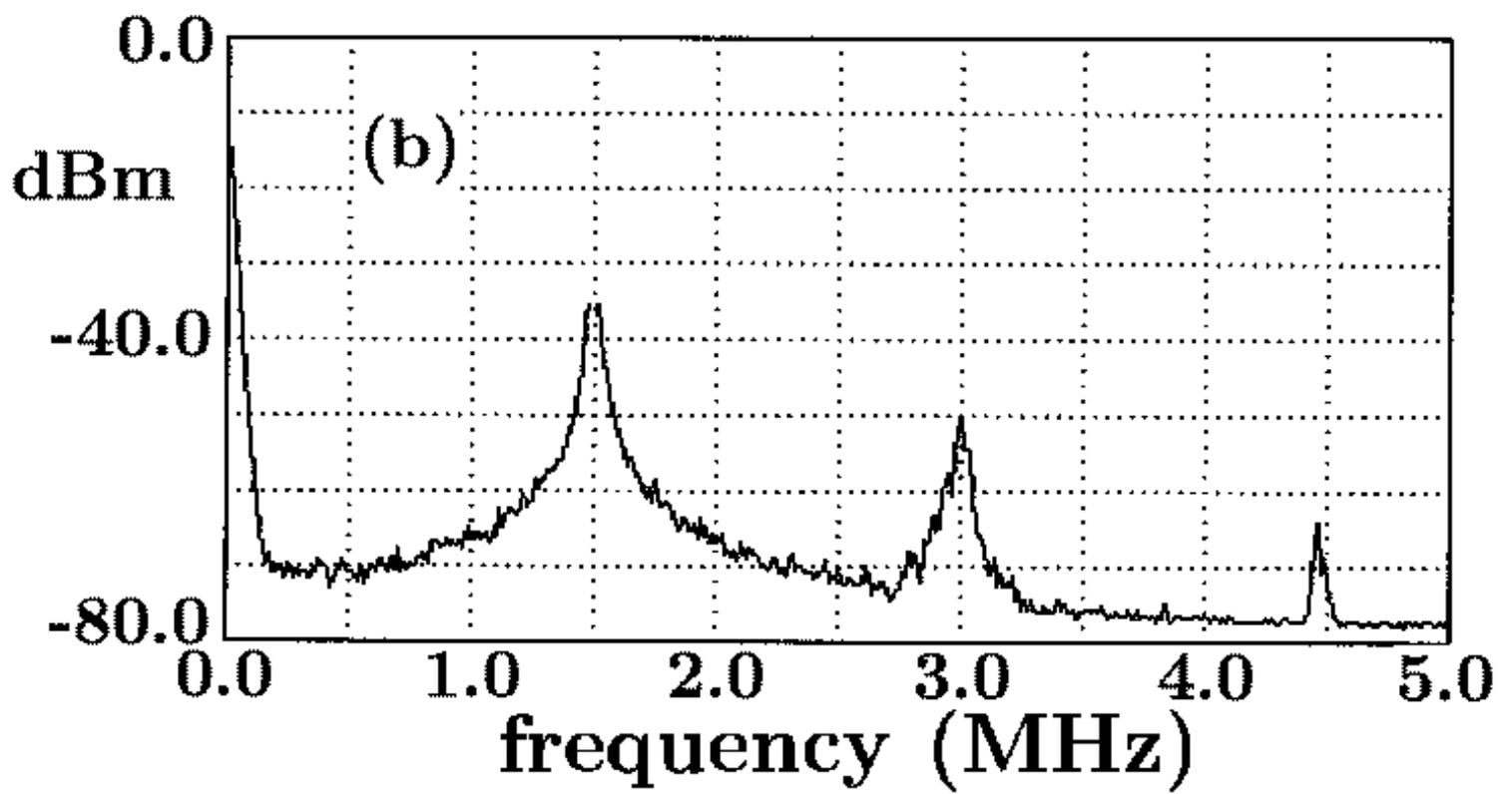
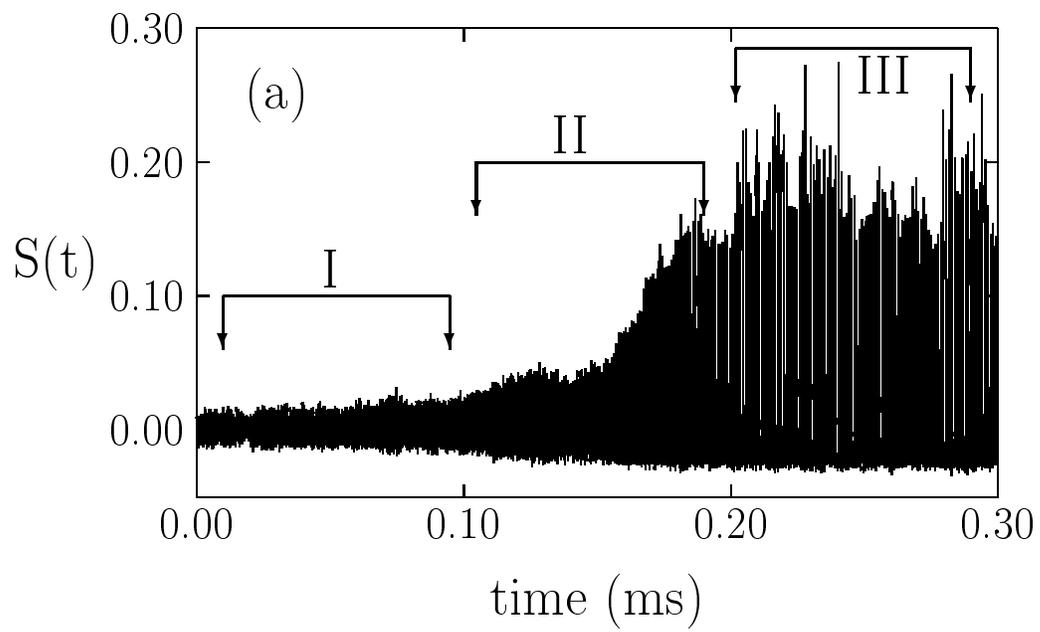
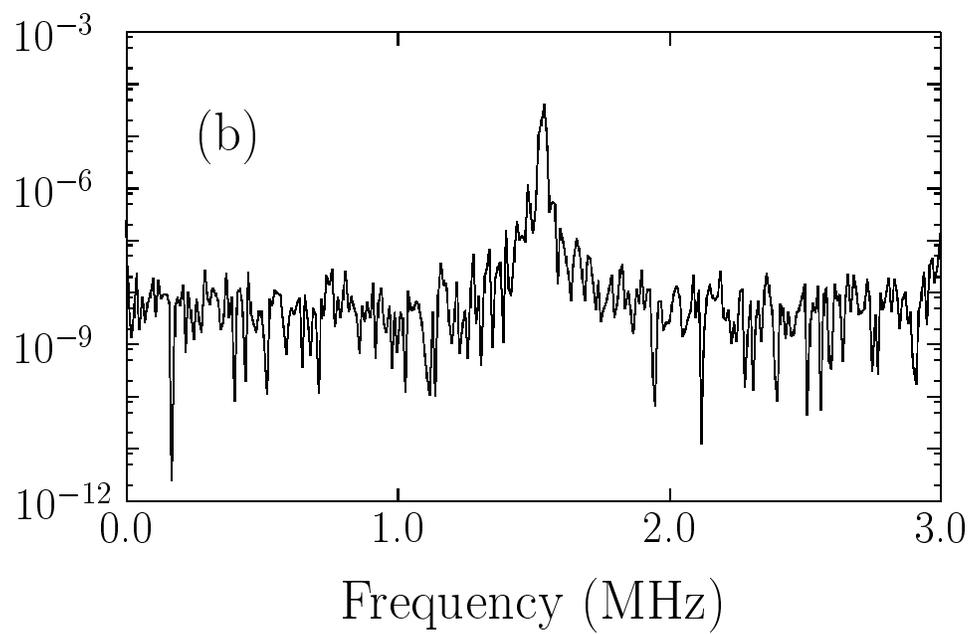
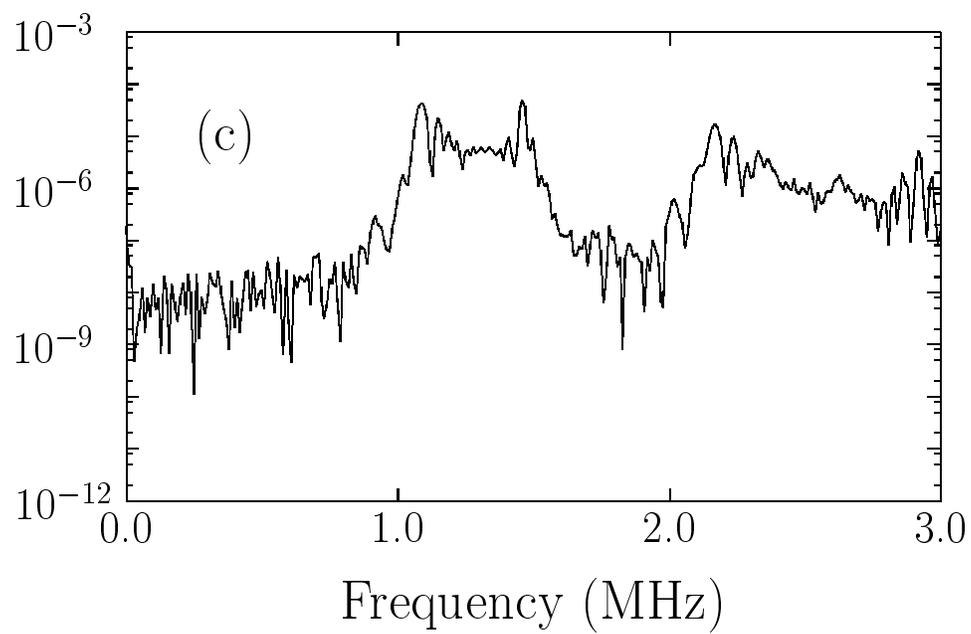
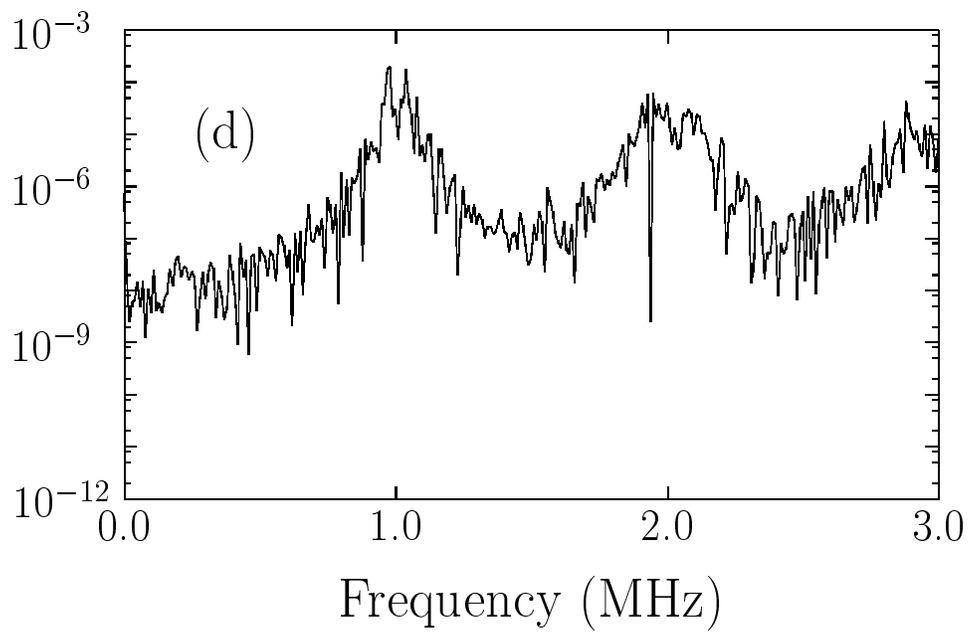
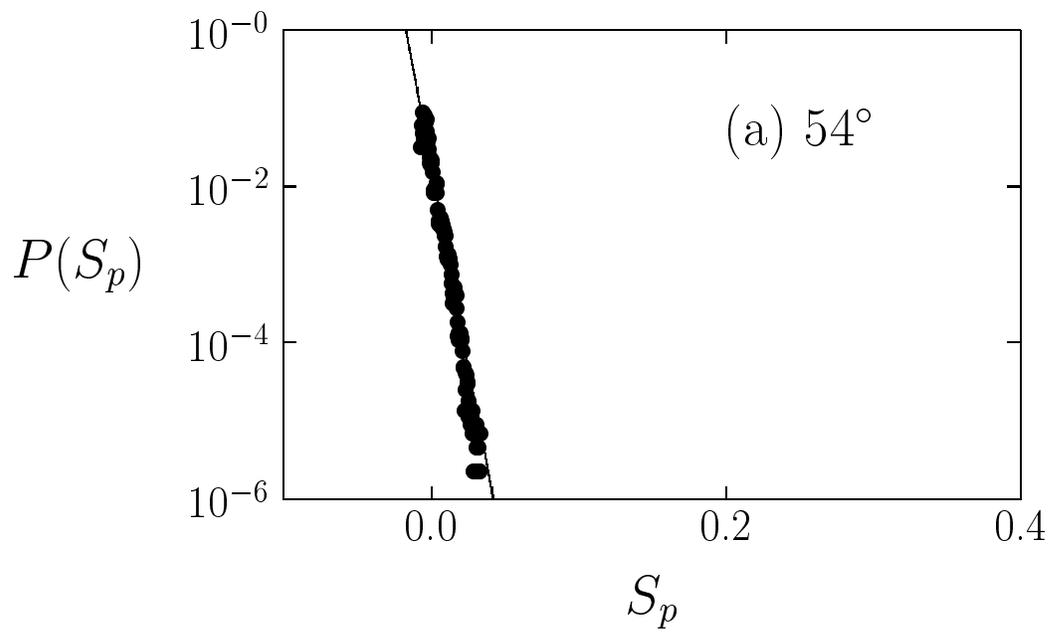
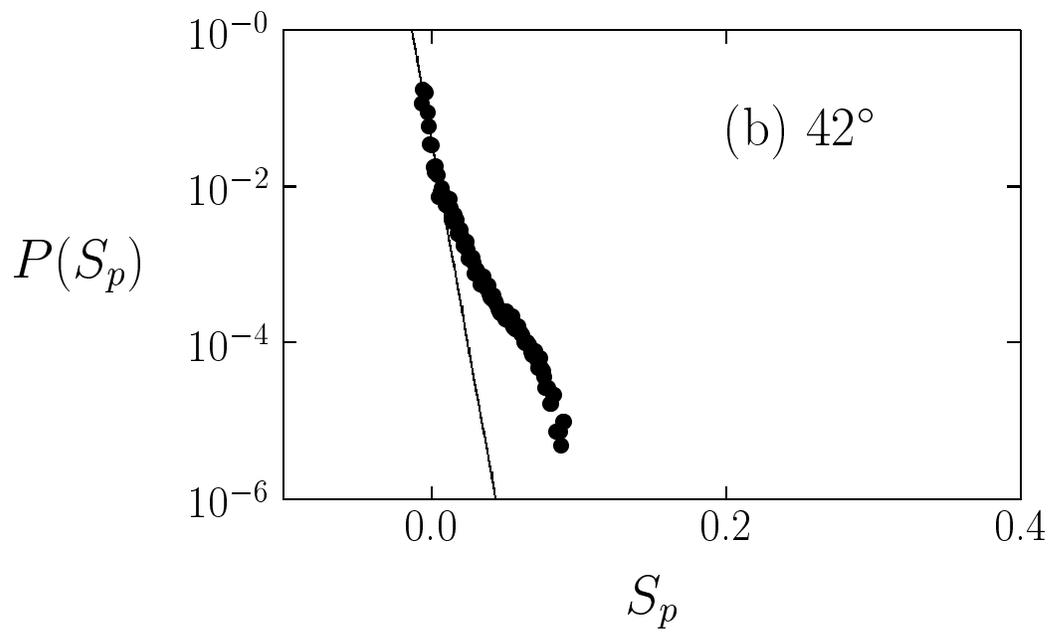
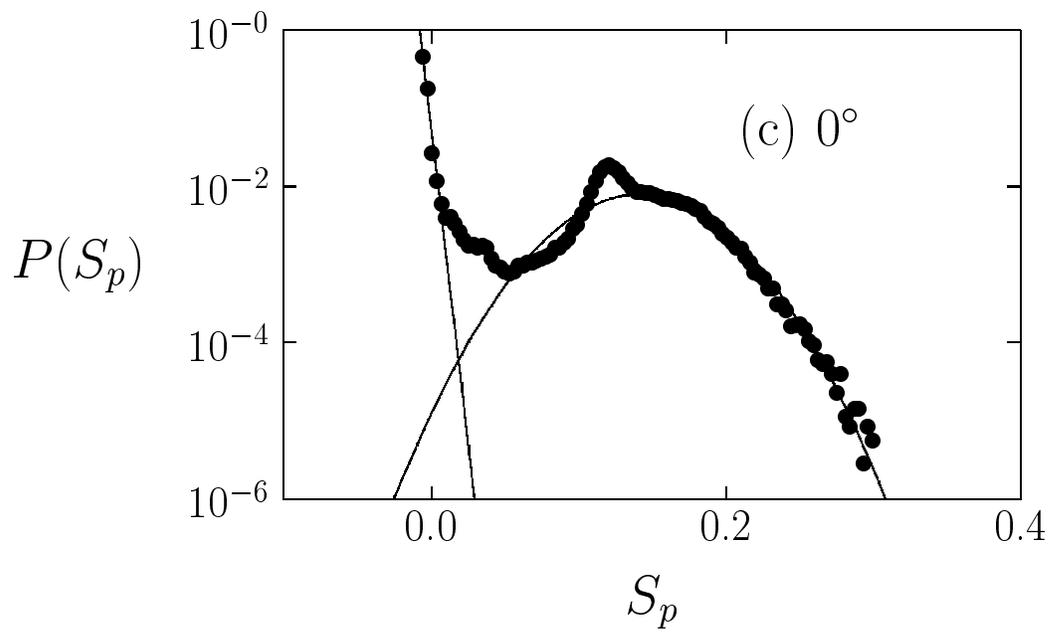
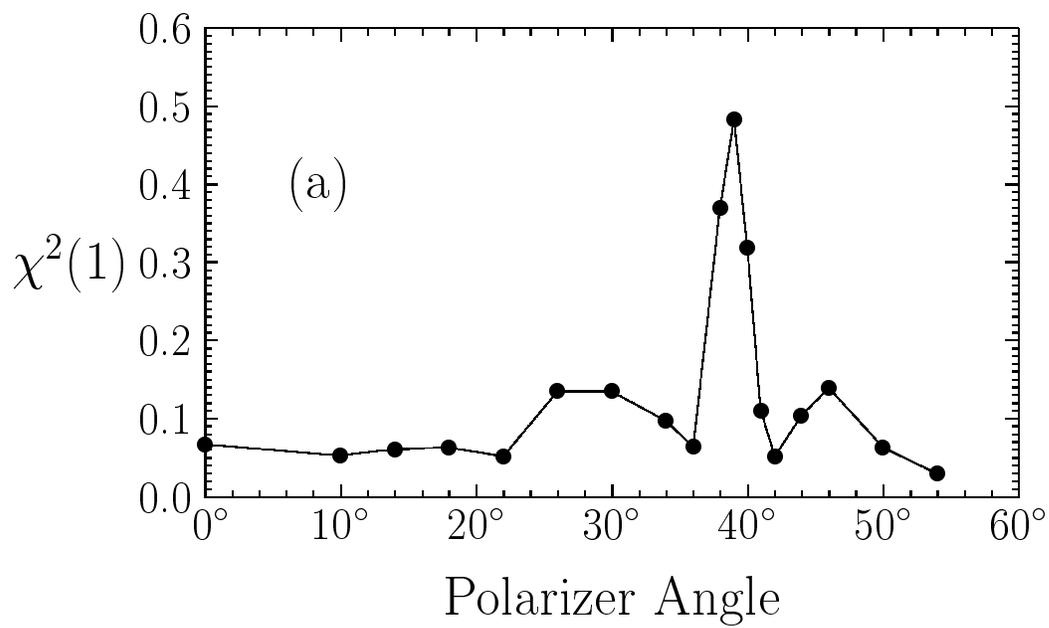
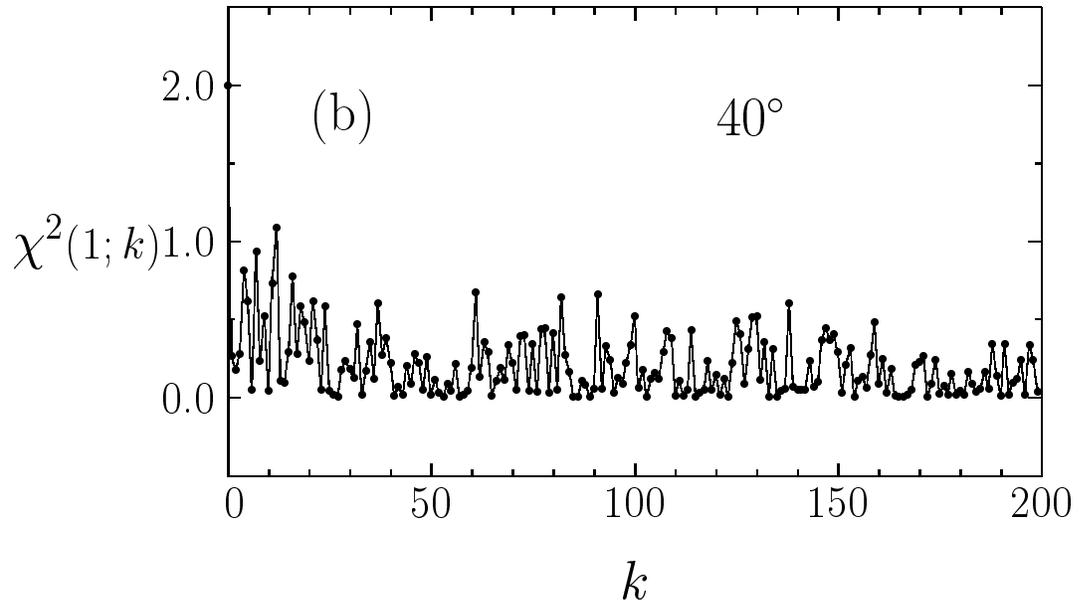
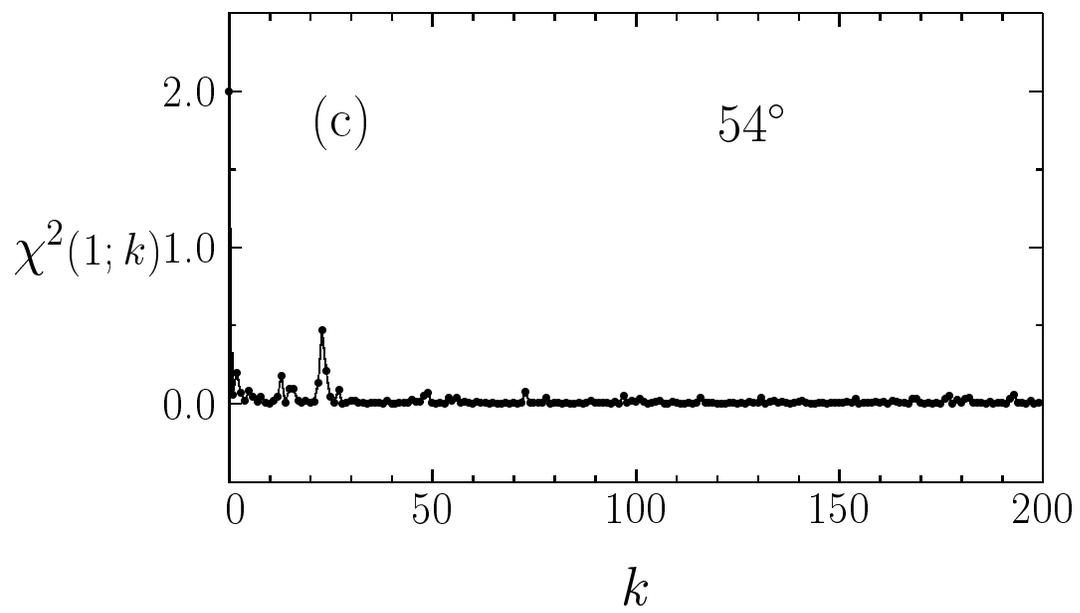
\end{document}